\documentstyle[psfig,preprint,aps]{revtex}

\tightenlines

\newcommand{\sla}{\!\!\!\!/ \,}

\def\beq{\begin{equation}}
\def\eeq{\end{equation}}
\def\bea{\begin{eqnarray}}
\def\eea{\end{eqnarray}}

\begin{document}
\hoffset-1cm
\draft

\title{Collisional Energy Loss of Fast Charged Particles in Relativistic 
Plasmas}

\author{Markus H. Thoma\footnote{Heisenberg fellow}}
\address{Theory Division, CERN, CH-1211 Geneva 23, Switzerland}

\date{\today}

\maketitle

\begin{abstract}

Following an argument by Kirzhnits we rederive an exact expression 
for the energy loss of a fast charged particle in a relativistic 
plasma using the quantum field theoretical language. We compare this result to 
perturbative calculations of the collisional energy loss of an energetic
electron or muon in an electron-positron plasma and of an energetic parton in 
the quark-gluon plasma.

\end{abstract} 

\bigskip

{\hspace*{1.1cm}Keywords: Energy loss; Relativistic plasmas; Thermal Field 
Theory}
\medskip
\pacs{PACS numbers: 52.60+h, 12.38.Mh, 11.10.Wx}

\narrowtext
\newpage

The energy loss of a fast charged particle in a medium
is a well studied subject \cite{Jackson}. Recently the energy loss
of energetic particles, such as leptons and partons, in relativistic plasmas
has attracted great interest. In relativistic heavy ion collisions the energy 
loss of a high energy quark or gluon coming from primary hard collisions
in the fireball may serve as a signature for the quark-gluon plasma
formation \cite{Thoma1}. In Supernovae explosions the energy loss of neutrinos,
having a weak charge,
in the plasma surrounding the stellar core might be an important mechanism
for triggering the explosion \cite{Raffelt}.

The total energy loss of a particle in a medium can be decomposed into a
collisional and a radiative contribution. While the first one originates 
from the energy transfer to the medium particles, the latter one is caused by
radiation from the fast particle. Here we want to consider only the 
collisional component. Whereas the radiative energy loss dominates in the
case of partons or charged leptons \cite{Baier}, the collisional one is
dominant for neutrinos in a Supernova plasma due to the small coupling
of the neutrinos to the medium. 

In quantum field theory the collisional energy loss per unit length
is defined as 
\cite{Braaten1}
\beq 
\frac{dE}{dx} = \frac{1}{v}\> \int d\Gamma \> \omega,
\label{eq1}
\eeq
where $v$ is the velocity of the incident particle with energy $E$
and $\omega =E-E'$ the energy
transfer to the medium. The interaction rate $\Gamma $ can be calculated 
either from the matrix element of the process responsible for the energy 
loss or equivalently from the imaginary part of the self energy of 
the particle with four momentum $P=(E,{\bf p})$, and mass $M$, ($p=|{\bf p}|$)
\cite{Braaten1}
\beq
\Gamma (E) = -\frac{1}{2E}\> [1-n_F(E)]\> tr[(P\sla +M)\, Im\, \Sigma(E,p)],
\label{eq2}
\eeq
where $n_F(E)=1/[\exp(E/T)+1]$ is the Fermi distribution in the case
of a fermion propagating through a plasma of temperature $T$. In the following
we restrict ourselves first to electrons or muons with high energies $E\gg T$
in an electron-positron plasma. Furthermore we assume first
only small momentum and energy transfers, $\omega$, $k\ll T$.
Assuming a one-loop approximation for $\Sigma $ but allowing for the 
most general photon propagator, indicated by the blob in Fig.1, we find
\cite{Braaten1}
\beq
\left ({\frac{dE}{dx}}\right )_{soft} 
= \frac{e^2}{2\pi v^2}\> \int_0^{k^*} dk k \> \int_{-vk}^{vk}
d\omega \> [1+n_B(\omega)]\> \left [\rho_l(\omega, k)+\left (v^2-\frac{\omega^2}
{k^2}\right )\rho_t(\omega, k)\right ],
\label{eq3}
\eeq
where $k^*\ll T$ is the separation scale, decomposing the energy loss into a 
soft and a hard part. 
$n_B(\omega)=1/[\exp(\omega/T)-1]$ is the Bose 
distribution
and $\rho_{l,t}$ are the spectral functions of the full photon propagator, 
defined as 
\beq
D_{l,t}(k_0,k) = \int_{-\infty}^{\infty} d\omega \> 
\frac{\rho_{l,t}(\omega ,k)}{k_0-\omega+i\varepsilon}.
\label{eq4}
\eeq
At finite temperature the photon propagator has only two independent
components \cite{Kapusta}, given in Coulomb gauge by the longitudinal 
and transverse propagators \cite{Thoma1} 
\bea
D_l(k_0,k) &=& \frac{1}{k^2-\Pi_l(k_0,k)+i\varepsilon},\nonumber \\
D_t(k_0,k) &=& \frac{1}{k_0^2-k^2-\Pi_t(k_0,k)+i\varepsilon},
\label{eq5}
\eea
where $\Pi_{l,t}$ are the longitudinal and transverse components of the
polarization tensor. It should be noted that the soft collisional energy loss, 
discussed here, follows according to (\ref{eq3})
only from the exchange of one dressed space-like ($\omega^2-k^2<0$)
photon from the particle to the medium according to linear response theory.
However, the medium particles may undergo further interactions.
The physical process corresponding to the imaginary part of the self energy
of Fig1. can be found by using cutting rules. An example is shown in Fig.2.
There is no diagram, where to or more photons are emitted from the
fast particle, as it is the case e.g. for bremsstrahlung. In the case of a
neutrino, however, diagrams containing two gauge boson lines are suppressed
anyway.

The spectral functions can be expressed by the imaginary
part of the photon propagator according to 
\beq
\rho_{l,t}(\omega ,k)=-\frac{1}{\pi}\> Im\, D_{l,t}(\omega ,k).
\label{eq6}
\eeq
For soft energy transfers $\omega \ll T$, as we assumed above, the
photon distribution can be expanded, leading to
\beq
1+n_B(\omega )\simeq \frac{T}{\omega} + \frac{1}{2}.
\label{eq7}
\eeq
Substituting (\ref{eq7}) into (\ref{eq3}) only the second term in (\ref{eq7})
contributes since the spectral functions are odd functions of $\omega $
\cite{Schulz}.

Alternatively the soft energy loss can also be derived from classical plasma 
physics arguments.
It follows from the induced electric field of the fast charged particle
in the plasma, which reacts on the incident particle by the Lorentz force,
causing the energy loss \cite{Ichimaru}. This process is known as the
Fermi density effect \cite{Jackson}. Introducing the dielectric
functions of the medium, the soft energy loss can be written as
\cite{Thoma2} 
\beq
\left ({\frac{dE}{dx}}\right )_{soft} 
= -\frac{e^2}{4\pi^2 v^2}\> \int_0^{k^*} dk k\> \int_{-vk}^{vk}
d\omega \omega\> \left [k^2 Im\, {\epsilon_l(\omega, k)}^{-1}
+\left (v^2-\frac{\omega^2}{k^2}\right )\> Im\, (\omega^2 
\epsilon_t(\omega, k)-k^2)^{-1}\right ].
\label{eq8}
\eeq
This expression is equivalent to (\ref{eq3}), since the dielectric functions
are related to the polarization tensor via \cite{Thoma2,Heinz,Mrow}
\bea
\epsilon_l(\omega,k) &=& 1-\frac{\Pi_l(\omega,k)}{k^2},\nonumber \\
\epsilon_t(\omega,k) &=& 1-\frac{\Pi_t(\omega,k)}{\omega^2}
\label{eq9}
\eea
and therefore also to the spectral functions, which are given by the
imaginary part of the propagators (\ref{eq5}), by
\bea
&& Im\, \epsilon_l(\omega, k)^{-1} = - \pi k^2 \rho_l(\omega, k), \nonumber \\
&& Im\, (\omega \epsilon_t(\omega ,k)-k^2)^{-1} = -\pi \rho_t(\omega,k),
\label{eq11}
\eea
where only the discontinuous part of the spectral functions coming from
the imaginary part of the polarization tensor contributes.

Now we want to derive an exact result for the soft collisional energy loss
using a generalized Kramers-Kronig relation and the asymptotic behavior of
the dielectric functions. For this purpose we introduce the 
response function of the medium
$R(k_0,k)=R_l(k_0,k)+R_t(k_0,k)$, given by \cite{Kirzhnits}
\bea
R_l(k_0,k) &=& -\frac{1}{\epsilon_l(k_0,k)},\nonumber \\
R_t(k_0,k) &=& \frac{k^2-k_0^2}{k^2-k_0^2\epsilon_t(k_0,k)}.
\label{eq12}
\eea
Using (\ref{eq5}) and (\ref{eq9}) we obtain
\beq
R(k_0,k)=-k^2\> D_l(k_0,k)+(k_0^2-k^2)\> D_t(k_0,k).
\label{eq13}
\eeq

Replacing the spectral functions in (\ref{eq3}) or the dielectric
functions in (\ref{eq8}) by the quantity $R$, making the substitution
$k\rightarrow q=\sqrt{k^2-\omega^2}$, i.e. introducing the magnitude
of the four momentum of the exchanged photon, and using $Im\, R(-\omega)
=-Im\, R(\omega )$ we find
\beq
\left ({\frac{dE}{dx}}\right )_{soft} = 
\frac{e^2}{2\pi^2}\> \int_0^{q^*} dq q\> \int_{0}^{\infty}
d\omega \omega\> \frac{Im\, R(\omega,\sqrt{q^2+\omega^2})}{q^2+\omega^2}.
\label{eq14}
\eeq
Here we restricted ourselves to ultrarelativistic particles, $v=1$,
and $q^*\ll T$. Eq. (\ref{eq14}) agrees with Ref.\cite{Kirzhnits},
if we replace there $Q^2$ by $e^2/4\pi$. 

The response function $R$ fulfills the following Kramers-Kronig
relation \cite{Kirzhnits}
\beq
R(k_0,k)=\tilde R + \frac{2}{\pi}\> \int_0^\infty d\omega \omega\>
\frac{Im\, R(\omega,k)}{\omega^2-k_0^2-i\varepsilon},
\label{eq15}
\eeq
which can be shown to be equivalent to the definition of the spectral functions
(\ref{eq4}), if we use $\rho_{l,t}(-\omega)=-\rho_{l,t}(\omega)$. 
Here $\tilde R=\lim_{k_0 \rightarrow \infty} R(k_0, k)
\sim \lim_{k_0 \rightarrow \infty} 1/k_0^2 =0$. 
The relation (\ref{eq15}) can be generalized to the so-called Leontovich
relation \cite{Leontovich} exploiting causality, from which
one obtains \cite{Kirzhnits,Dolgov}
\beq
R(k_0,\sqrt{k^2+k_0^2})=R_\infty + \frac{2}{\pi}\> \int_0^\infty 
d\omega \omega\> \frac{Im\, R(\omega,\sqrt{k^2+\omega^2})}
{\omega^2-k_0^2-i\varepsilon},
\label{eq16}
\eeq
where $R_\infty=\lim_{k_0 \rightarrow \infty} R(k_0,\sqrt{k^2+k_0^2})$.

The $\omega $-integral 
\beq
I=\frac{2}{\pi}\>\int_{0}^{\infty}
d\omega \omega\> \frac{Im\, R(\omega,\sqrt{q^2+\omega^2})}{q^2+\omega^2}
\label{eq17}
\eeq 
appearing in  the energy loss (\ref{eq14}) agrees with the 
integral on the right hand side of the Leontovich relation, if we
replace $\omega $ by $iq$ and $\sqrt{k^2+\omega^2}$ by 0, i.e. $k^2=q^2$,
in (\ref{eq16}). Therefore we can write \cite{Kirzhnits}
\beq
I=R(iq,0)-R_\infty.
\label{eq18}
\eeq
Since the longitudinal and the transverse dielectric functions are identical
at zero momentum \cite{Lifshitz}, $\epsilon_l(k_0,0)=\epsilon_t(k_0,0)$, we
have $R(iq,0)=0$. $R_\infty$ is related to 
the high frequency and momentum limit of the dielectric 
functions, which agrees with the vacuum result $\epsilon_{l}=1$.
For the transverse part we have to consider corrections to
the vacuum value. From the second equation of (\ref{eq9}) we get
\beq
\lim_{k_0\rightarrow \infty} \epsilon_t(k_0, \sqrt{q^2+k_0^2})
=1-\frac{\omega_0^2}{k_0^2},
\label{eq19}
\eeq
where 
\beq
\omega_0^2\equiv \lim_{k_0\rightarrow \infty}\Pi_t(k_0,\sqrt{q^2+k_0^2})
\label{eq20}
\eeq
Using the Kramers-Kronig relation for the transverse dielectric functions
it can be shown \cite{Kirzhnits}, that $\omega_0$ is independent of $q$.
It can be considered as the effective thermal mass of the transverse 
high frequency plasma excitations, which is given by $\omega_0^2=e^2 n \langle 
1/\Omega \rangle$ in the relativistic limit \cite{Kirzhnits}. Here $n$ is the 
number density of the medium and $\Omega $ the energy of the plasma particles. 
In the non-relativistic limit $\omega_0$ is identical to the plasma frequency
\cite{Kirzhnits}.

Using (\ref{eq12}) together with the high frequency and momentum limit of
the dielectric functions we get
\beq
I=-R_\infty=\frac{\omega_0^2}{q^2+\omega_0^2}.
\label{eq21}
\eeq
Combining this result for $I$ with (\ref{eq14}) we end up with 
\beq
\left ({\frac{dE}{dx}}\right )_{soft} = \frac{e^2}{4\pi}\> \omega_0^2 \ln 
\frac{q^*}{\omega_0},
\label{eq22}
\eeq
where we assumed $q^*\gg \omega_0$. 

The unknown parameter $\omega_0$ following from the full transverse
polarization tensor serves as an infrared cutoff for the photon exchange.
Since the total collisional energy loss has to be independent of the 
arbitrary separation scale $q^*$, the hard part has to assume the form
\beq
\left ({\frac{dE}{dx}}\right )_{hard} = \frac{e^2}{4\pi}\> \omega_0^2 \ln 
\frac{q_{max}}{q^*},
\label{eq23}
\eeq
where $q_{max}$ is proportional to the maximum energy transfer,
i.e. $q_{max} \sim \sqrt{ET}$ in the relativistic limit
$E\gg \Omega$ \cite{Kirzhnits}, which we have considered here.
Note that the hard contribution to the energy loss contains besides $t$-channel
diagrams, as the one in Fig.2, also $s$- and $u$-channel ones, which,
however, do not contribute to the leading logarithm.

Hence we obtained a very simple 
expression 
\beq
{\frac{dE}{dx}} = \frac{e^2}{4\pi}\> \omega_0^2 \ln 
\frac{q_{max}}{\omega_0},
\label{eq24}
\eeq
for the exact result of the collisional energy loss,
independent of any approximation to the full photon propagator
or the dielectric functions of the medium, respectively. To logarithmic 
accuracy the final result just depends on the parameter $\omega_0$.

As an example we consider the high temperature limit of the energy loss.
There it can be calculated to leading order perturbation theory using
the Hard Thermal Loop (HTL) 
resummation technique \cite{Braaten2}. Computing the soft
energy loss using the HTL resummed photon propagator in Fig.1 and calculating
the hard part from the tree level scattering matrix elements
one finds in the limit $v=1$ \cite{Braaten1}
\beq
{\frac{dE}{dx}} = \frac{e^2}{4\pi}\> \omega_0^2 \left (\ln 
\frac{\sqrt{ET}}{\omega_0}+0.120\right ),
\label{eq25}
\eeq
where $\omega_0^2 = 3m_\gamma^2/2$. The thermal photon mass $m_\gamma$,
which is equivalent to the plasma frequency, is given by $m_\gamma =eT/3$.
Indeed $\omega_0$ is given by the high frequency and momentum
limit (\ref{eq20}) of the transverse HTL polarization tensor \cite{Pisarski}
\beq
\Pi_t^{HTL}(k_0,k)=\frac{3}{2}\> m_\gamma^2 \frac{\omega^2}{k^2}\>
\left [1-\left(1-\frac{k^2}{k_0^2}\right )\> \frac{k_0}{2k}\> \ln
\frac{k_0+k}{k_0-k}\right ]
\label{eq26}
\eeq
confirming the general result (\ref{eq24}) in the HTL limit. Also
$\omega_0^2=e^2n\langle 1/\Omega\rangle$ holds for the HTL case,
where \cite{Carrington}
\beq
m_\gamma^2=\frac{4e^2}{3\pi^2}\> \int_0^\infty dk k\> n_F(k),
\label{eq26a}
\eeq
because
\beq
n=4\int \frac{d^3k}{(2\pi)^3}\> n_F(k), \; \; \; \; \; \; \; 
\left \langle \frac{1}{\Omega}\right \rangle =\frac{\int \frac{d^3k}{(2\pi)^3}
\> \frac{1}{k}\> n_F(k)}{\int \frac{d^3k}{(2\pi)^3}\> n_F(k)}. 
\label{eq26b}
\eeq

In the case of the collisional energy loss of quarks or gluons in a 
quark-gluon plasma we simply have to replace the factor
$e^2$ in (\ref{eq24}) by $C_Fg^2=4g^2/3$ for quarks and by
$C_Ag^2=3g^2$ for gluons, where $g$ is the strong coupling constant.
Furthermore $\omega_0^2$ is now the high frequency and momentum limit
of the transverse gluon polarization tensor, which is given by
$3m_g/2$, where the effective gluon mass reads $m_g^2=g^2T^2(1+n_f/6)/3$
in a QGP containing $n_f$ thermalized quark flavors. Using the HTL
limit for $\omega_0$ in (\ref{eq24}) we reproduce again the result
obtained from an explicit HTL resummed calculation within the logarithmic
approximation\cite{Braaten3}. Another application of (\ref{eq24}) has been
discussed in Ref.\cite{Kirzhnits2} in connection with the neutrino
energy loss in matter.

Summarizing, we have shown that different definitions of the collisional 
energy loss, based either on quantum field theory or on plasma physics,
are equivalent. Translating the arguments, based on a generalized
Kramers-Kronig relation, given by Kirzhnits \cite{Kirzhnits}
to a quantum field theoretical language, using self energies, propagators and 
their spectral functions, we gave an exact result for the collisional energy 
loss (\ref{eq24}). Within the logarithmic approximation 
it contains the effective 
mass of the high frequency transverse plasma mode as the only parameter.
Assuming that this mass is given approximately by the high temperature
result, we obtain an simple estimate 
for the collisional energy loss for energetic
electrons and muons in a QED plasma and for partons in the quark-gluon plasma.
Finally we showed that the perturbative result for the collisional energy loss,
obtained within the HTL resummation method, is in agreement with the general 
result found by Kirzhnits.

\newpage

\centerline{\bf ACKNOWLEDGMENTS}
\vspace*{0.5cm}
The author is grateful to G. Raffelt for drawing his attention to the
paper by D.A. Kirzhnits and for helpful discussions and to the 
Max-Planck-Institut f\"ur Physik (Werner-Heisenberg-Institut) for their 
hospitality.

\begin{figure}

\centerline{\psfig{figure=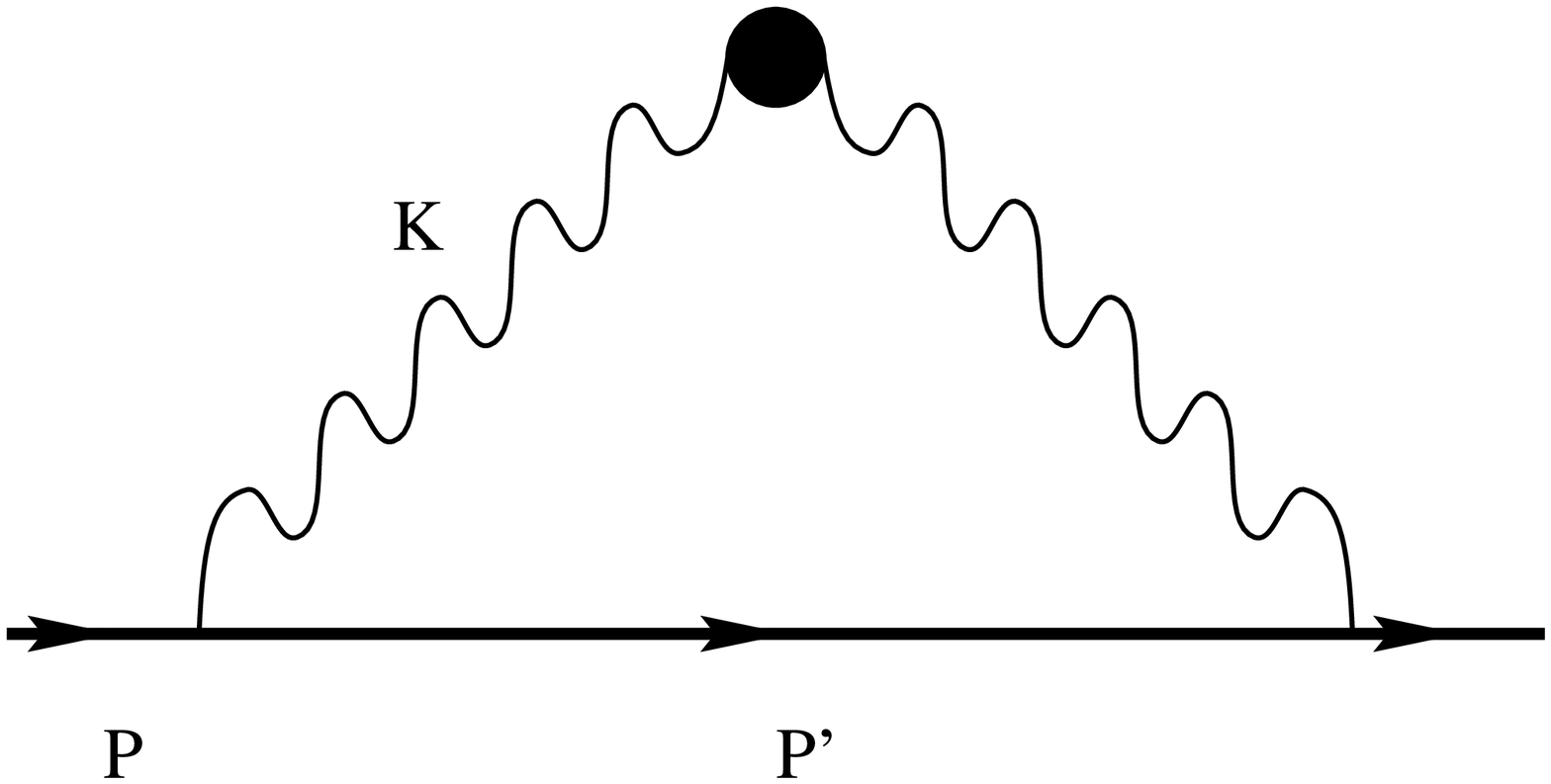,width=10cm}}
\caption{Self energy of a fast fermion containing the full gauge boson 
propagator}

\end{figure}

\begin{figure}

\centerline{\psfig{figure=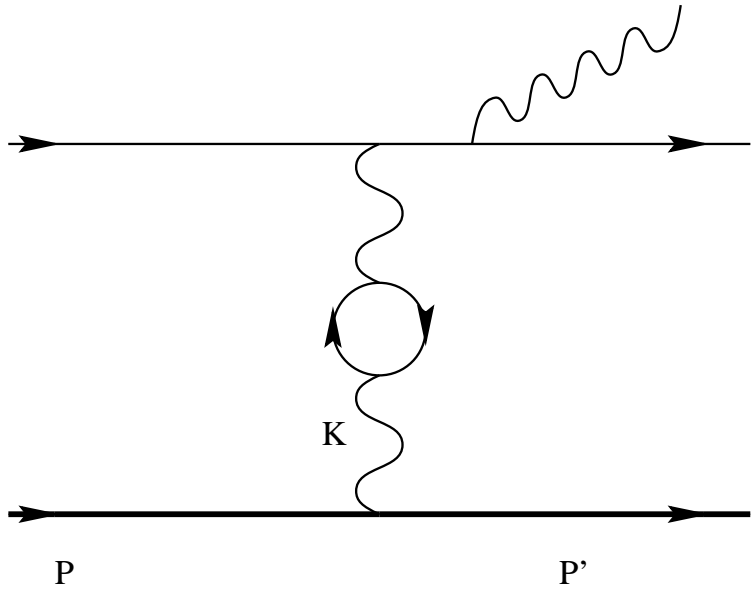,width=10cm}}
\caption{Example for a scattering diagram related to the imaginary part 
of the diagram in Fig.1}

\end{figure}

\end{document}